\documentclass[pra,superscriptaddress,showpacs,twocolumn]{revtex4-1}

\usepackage{amsmath,amsfonts,amsthm,amssymb}
\usepackage{graphicx}
\usepackage{graphics}
\usepackage{braket}
\usepackage{setspace}
\usepackage{color}
\usepackage{ulem}
\usepackage{placeins}

\DeclareMathOperator{\Tr}{Tr}

\newcommand{\C}{{\cal C}}
\newcommand{\oldvec}[1]{#1}	% remove any functionality from 'oldvec'
\begin{document}

\title{Multipartite entanglement detection with minimal effort}

\author{Lukas~Knips}
\affiliation{Max-Planck-Institut f\"ur Quantenoptik, Hans-Kopfermann-Strasse 1, D-85748 Garching, Germany}
\affiliation{Fakult\"at f\"ur Physik, Ludwig-Maximilians-Universit\"at, D-80797 M\"unchen, Germany}

\author{Christian~Schwemmer}
\affiliation{Max-Planck-Institut f\"ur Quantenoptik, Hans-Kopfermann-Strasse 1, D-85748 Garching, Germany}
\affiliation{Fakult\"at f\"ur Physik, Ludwig-Maximilians-Universit\"at, D-80797 M\"unchen, Germany}

\author{Nico~Klein}
\affiliation{Max-Planck-Institut f\"ur Quantenoptik, Hans-Kopfermann-Strasse 1, D-85748 Garching, Germany}
\affiliation{Fakult\"at f\"ur Physik, Ludwig-Maximilians-Universit\"at, D-80797 M\"unchen, Germany}

\author{Marcin~Wie\'{s}niak}
\affiliation{Institute of Theoretical Physics and Astrophysics, University of Gda\'nsk, PL-80-952 Gda\'nsk, Poland}

\author{Harald~Weinfurter}
\affiliation{Max-Planck-Institut f\"ur Quantenoptik, Hans-Kopfermann-Strasse 1, D-85748 Garching, Germany}
\affiliation{Fakult\"at f\"ur Physik, Ludwig-Maximilians-Universit\"at, D-80797 M\"unchen, Germany}

\begin{abstract}
Certifying entanglement of a multipartite state is generally considered as a demanding task. 
Since an $N$ qubit state is parametrized by $4^{N}-1$ real numbers, one might naively expect that the measurement effort of generic entanglement detection also scales 
exponentially with $N$. 
Here, we introduce a general scheme to construct efficient witnesses requiring a constant number of measurements independent of the number of qubits for states like, e.g., Greenberger-Horne-Zeilinger states, cluster states and Dicke states.
For four qubits, we apply this novel method to experimental realizations of the aforementioned states and prove genuine four-partite entanglement with two measurement settings only.
\end{abstract}
% 50 word abstract:
% Although for some specific quantum states, operators exist that witness genuine multipartite entanglement with only a few measurements, a constructive and simultaneously efficient method is missing. We here introduce such a scheme, tremendously reducing the experimental effort for many states, where no efficient witness is known.

\maketitle

\textit{Introduction.---}Entanglement is a fascinating feature of strictly quantum nature.
It was first studied for the bipartite case~\cite{EPR,Bell} and has already been applied for first quantum communication tasks like quantum cryptography and quantum teleportation~\cite{PanRev}.
The generalization to multipartite entanglement comes with a whole new set of features providing, relative to separable states, information processing advantages for quantum computation and simulation or for quantum metrology.
It is thus crucial to have tools at hand which allow to identify \textit{genuinely multipartite entangled} states ~\cite{BZPZ, HorodeckiRev, GuehneTothRev}.

Proving genuine multiparty entanglement is in general a complex task.
Full quantum state tomography  (QST) can be used for detecting and even for quantifying entanglement, but requires the determination of exponentially many parameters.
Even when using simplified procedures~\cite{PIT,MPS,LORANK}, the effort is still significant. 
Thus, it was the goal to find a direct measurement procedure for witnessing entanglement~\cite{Horodecki96,Lewenstein,MEW,GuehneTothRev}.
The only systematic method known today for constructing entanglement witnesses uses the fidelity relative to a chosen reference state. 
However, depending on the state, this as well leads to a rapidly increasing number of measurements required to infer the fidelity. 
Remarkably, for cluster and Greenberger-Horne-Zeilinger (GHZ) states, witnesses have been found incidentally which require only two measurements for any number of qubits~\cite{OtGe}.

In this letter, we introduce a constructive scheme to derive efficient multipartite entanglement witnesses, i.e., witnesses which can be evaluated from only a very small number of measurements. 
Our scheme employs basic properties of operators and their expectation values to construct witnesses for many relevant quantum states which require only two measurement settings, \textit{independent} of the number of qubits.
We demonstrate how to derive these efficient entanglement criteria for several of the most prominent quantum states, encompassing GHZ and cluster states, Dicke and W states, and the multipartite singlet state.

Every quantum mechanical $N$-qubit state $\rho$ is uniquely described by its correlation tensor $T$,
\begin{equation}
\rho = \frac{1}{2^N} \sum_{\oldvec{j}\in{\cal I}} T_{\oldvec{j}} \sigma_{\oldvec{j}},
\end{equation}
where the set ${\cal I}=\{0\dots00, 0\dots01, \dots, 3\dots33\}$ labels all indices $\oldvec{j}=\left(j_1 \dots j_N\right)$, $j_i\in\{0,1,2,3\}$ of the correlation tensor with
$\sigma_{\oldvec{j}}=\sigma_{j_1}\otimes\dots\otimes\sigma_{j_N}$ and with Pauli matrices $\sigma_0$, $\sigma_1$, $\sigma_2$, and $\sigma_3$.
The correlation tensor elements (for short called \textit{correlations}) are given
by $T_{\oldvec{j}} = \langle \sigma_{\oldvec{j}}\rangle = \Tr[\rho \sigma_{\oldvec{j}}]$. 
Since the eigenvalues of $\sigma_{\oldvec{j}}$ are $\pm 1$, the correlations are constrained to lie in the interval $[-1,1]$ and consequently $T_{\oldvec{j}}^2\leq 1$.
These constraints, together with the physicality condition $\rho\geq0$ imply various bounds on the summed squares of correlations, which are helpful for the construction of efficient witness operators. 
Consider a set of $n$ pairwise commuting operators $\{\sigma_{\oldvec{j}}: \oldvec{j}\in{\cal C} \subset{\cal I} \}$. 
These operators have common eigenstates, for which $ T_j = \pm 1$ holds.
Consequently, the sum of squared correlations is bounded by $\sum_{\oldvec{j} \in {\cal C}}T_{\oldvec{j}}^2\leq n$.
On the contrary, for a set of pairwise anticommuting operators, e.g., $\{\sigma_{\oldvec{j}}: \oldvec{j}\in{\cal A} \subset{\cal I} \}$, the threshold is~\cite{WIEMAR}
\begin{equation}
\sum\limits_{\oldvec{j} \in \cal{A}}T_{\oldvec{j}}^2\leq 1,
\label{eq:anticommutation_bound}
\end{equation}
establishing a complementarity relation between the correlations~\cite{Kurzynski}. 

\textit{Separability.---}
Consider the bipartition (\textit{cut}) ${\cal B}=A|B$ of a multipartite quantum system into parts $A$ and $B$.
Two operators given by $\sigma_{ab}=\sigma_a \otimes \sigma_b$ and $\sigma_{a'  b'}=\sigma_{a'} \otimes \sigma_{ b'}$ \textit{anticommute with respect to the bipartition ${\cal B}$} if $\{\sigma_a, \sigma_{a'}\}=0$ or $\{\sigma_b, \sigma_{b'}\}=0$, i.e., if they locally anticommute on $A$ or on $B$.
We call this property \textit{cut-anticommutativity} or, more specifically, $A|B$-anticommutativity.
Since for states separable with respect to ${\cal B}$ the correlation tensor factorizes, $T_{ab}=T_a T_b$, these states fulfill
\begin{equation}
 T_{ab}^2+T_{a' b'}^2\leq_{\mathop{\rm SEP}\limits_{\cal B}} 1,
 \label{eq:complementarity}
\end{equation}
see~\cite{WIEMAR}.
However, cut-anticommuting operators can also commute, i.e., $[\sigma_{ab},\sigma_{a' b'}]=0$, allowing the common (entangled) eigenstates of $\sigma_{ab}$ and $\sigma_{a' b'}$ to exhibit $T_{ab}^2+T_{a' b'}^2 > 1$. 
Therefore, violation of Eq.~(\ref{eq:complementarity}) rules out separability with respect to cut ${\cal B}$.

\textit{Testing entanglement.---}
To prove genuine multipartite entanglement of a state, Eq.~(\ref{eq:complementarity}) has to be violated for every possible bipartition. 
One starts with a list $\{\sigma_j\}$ of all operators with nonvanishing expectation value, $T_j\neq0$ (all non-vanishing correlations). 
For the construction of the efficient entanglement criterion for a bipartition ${\cal B}$, one then chooses from that list two operators which are mutually commuting, but also cut-anticommuting relative to the bipartition $A|B$.
One repeats this, until all bipartitions are tested.

The scheme becomes highly efficient if the correlation values of several $\sigma_j$ can be obtained from the same measurement setting. 
In detail, this means that one makes use of the observation that from a single measurement setting ${\cal M}_k$ with $k=(k_1,k_2,...,k_N)$ and $k_i \in \{1,2,3\} $ labeling the local Pauli measurements, all $2^N$ correlations $T_j$ with $j \in \{(0,0,...,0),(0,0,...,k_N),...,(k_1,k_2,...,k_N) \} $ can be inferred.
Depending on the symmetry of the state, two measurement settings 
can suffice to prove genuine multipartite entanglement if one finds for each bipartition operators in the set that are commuting, but cut-anticommuting for the given bipartition. 

\textit{Combined entanglement witness.---}
Combining the above criteria into a single witness facilitates the practical application (only a single value has to be calculated), though at the expense of a lower sensitivity, i.e., a reduced robustness against (white) noise.
To optimize the sensitivity, a weighted sum is used,
\begin{equation}
{\cal W}=\frac{1}{G_0} \sum\limits_{\oldvec{j} \in {\cal S}} v_{\oldvec{j}} T_{\oldvec{j}}^2 \leq_{\rm BISEP} \frac{G}{G_0},
\label{eq:generalCriterion}
\end{equation}
where ${\cal S}\subset{\cal I}$ labels the set of correlations that can be determined by the given set of measurements and where $\leq_{\rm BISEP}$ denotes that the inequality is valid for all biseparable states. 
The weights $v_{\oldvec{j}}$ and the (normalization) constants $G$ ($G_0$) are determined as follows:\newline
\textit{i)} Depict the operators defined by ${\cal S}$ as vertices of a graph (\textit{anticommutativity graph}).\\
\textit{ii)} Assign weights $v_{\oldvec{j}}>0$ to the vertices.\\
\textit{iii)} Choose bipartition ${\cal B}_r$ and connect all vertices for which the corresponding operators cut-anticommute by edges. (If all operators indexed by ${\cal S}$ mutually commute, no edges will occur.)
Distribute values $c^{(m)}_{\oldvec{j}} = \{0,1\}$ among vertices under the constraint that any two `$1$'s are not connected by an edge and calculate 
for each of the $m$ possible distributions of `$1$'s the sum ${G}_r^{(m)}=\sum_{\oldvec{j} \in {\cal S}} c_{\oldvec{j}}^{(m)} v_{\oldvec{j}}$. 
The case of no partition will be labeled by $r=0$.
Repeat step \textit{iii)} for all bipartitions ${\cal B}_{r}$.\\
\textit{iv)} Every choice of weights $v_{\oldvec{j}}$ in Eq.~(\ref{eq:generalCriterion}) defines a witness 
with ${G}=\max_{r>0,m} {G}_r^{(m)}$ and $G_0=\max_{m} {G}^{(m)}_0$. 
The ratio ${G}/G_0$ determines the noise robustness of the criterion. 
To optimize the witness in terms of its noise robustness, one has to choose the weights $v_{\oldvec{j}}$ according to $\arg \min_{\{v_{\oldvec{j}}\}} {G}/G_0$.

\begin{center}
\begin{table}[b]
\caption{All nonvanishing correlations of the four-qubit GHZ state. 
The correlations colored in blue can be inferred from the measurement setting ${\cal M}_{3333}$ and the correlation colored in red is
obtained from the setting ${\cal M}_{1221}$.
}
\vspace{0.2cm}
\begin{tabular}[b]{|c |r |c| c|r |c|c|r|c|c|r|} 
\hline \hline 
$T_{0000}$ 	& $1$ & &
$\textcolor{blue}{\boldsymbol{T_{0033}}}$ 	& $1$ & &
$\textcolor{blue}{\boldsymbol{T_{0303}}}$ 	& $1$ & &
$\textcolor{blue}{\boldsymbol{T_{0330}}}$ 	& $1$ \\
$\textcolor{blue}{\boldsymbol{T_{3003}}}$ 	& $1$ & &
$\textcolor{blue}{\boldsymbol{T_{3030}}}$ 	& $1$ & &
$\textcolor{blue}{\boldsymbol{T_{3300}}}$ 	& $1$ & &
$\textcolor{blue}{\boldsymbol{T_{3333}}}$ 	& $1$ \\
$T_{2112}$ 	& $-1$ & &
$T_{2121}$ 	& $-1$ & &
$T_{2211}$ 	& $-1$ & &
$T_{2222}$ 	& $1$ \\
$T_{1111}$ 	& $1$ & &
$T_{1122}$ 	& $-1$ & &
$T_{1212}$ 	& $-1$ & &
$\textcolor{red}{\boldsymbol{T_{1221}}}$ 	& $-1$ \\
\hline \hline
\end{tabular}
\label{tab:corrsGHZ}
\end{table}
\end{center}

\textit{Example.---}Let us consider the four-party GHZ state $\frac{1}{\sqrt{2}}(|0000\rangle+|1111\rangle)$, whose nonvanishing correlations are listed in Tab.~\ref{tab:corrsGHZ}. 
As one can see, the measurement of the single setting ${\cal M}_{3333}$ provides $7$ correlations with squared value $1$ (marked blue). 
Since the operators of these correlations exhibit the same cut-anticommutation relation with any operator corresponding to the other $8$ correlations of Tab.~\ref{tab:corrsGHZ}, the second measurement can be chosen arbitrarily out of those remaining $8$. 
For example, the choice ${\cal M}_{1221}$ for the second measurement setting results in the set of operators
$\{\sigma_{3333}$, $\sigma_{3300}$, $\sigma_{0033}$, $\sigma_{3003}$, 
$\sigma_{0330}$, $\sigma_{3030}$, $\sigma_{0303}$, $\sigma_{1221}\}$, i.e., ${\cal S}=\{3333, 3300, \dots, 1221\}$.

States that are, e.g., $A|BCD$-separable fulfill, according to Eq.~(\ref{eq:complementarity}),
\begin{align}
T_{3333}^2+T_{1221}^2\leq_{\mathop{\rm SEP}\limits_{A|BCD}}1.
\label{eq:GHZ_first_A_BCD}
\end{align}
Since $\sigma_{1221}$ not only $A|BCD$-anticommutes with $\sigma_{3333}$, but also with $\sigma_{3030}$, $\sigma_{3003}$, $\sigma_{3300}$ from our list, a natural choice is to average over the expectation values of those $4$ possibilities.
Nonseparability against the partition $A|BCD$ can then be detected with
\begin{align}
{\cal W}^{\rm GHZ}_{A|BCD}=\frac{1}{2}\Big[{\textstyle \frac{1}{4}}\left(T_{3030}^2+T_{3003}^2+T_{3300}^2+T_{3333}^2\right)\nonumber\\
+T_{1221}^2\Big]\leq_{\mathop{\rm SEP}\limits_{A|BCD}}\frac{1}{2},
\label{eq:GHZ_A_BCD}
\end{align}
where the additional normalization constant of $1/2$ is introduced to ensure that ${\cal W}^{\rm GHZ}_{A|BCD}=1$ holds for the ideal GHZ state, where all squared expectation values are one. 
The criteria for the remaining six bipartitions are derived analogously. 
For the list of criteria for the four-qubit Dicke, singlet, and W state see the supplemental material (SM)~\cite{SM}. \\

\begin{center}
\begin{figure}
\includegraphics[width=0.45\textwidth]{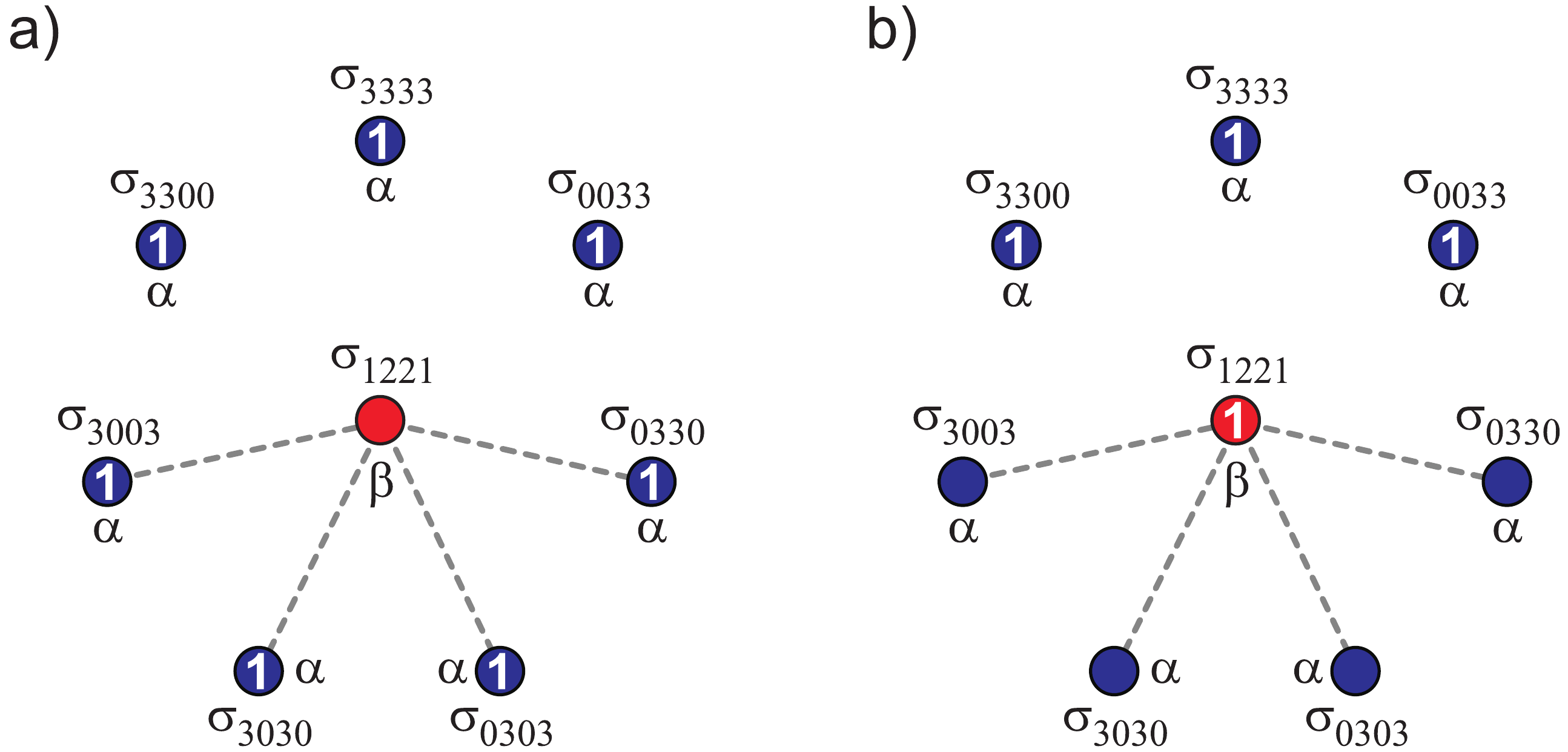}
\caption{The operators used to construct the witness ${\cal W}^{\rm GHZ}$, using the same color code as in Tab.~\ref{tab:corrsGHZ}. 
As an example, the cut-anticommutation relations for the cut $AB|CD$ are indicated by dashed lines.
One realizes that for each bipartition four of the seven operators obtained from the measurement of setting ${\cal M}_{3333}$ cut-anticommute with $\sigma_{1221}$. 
Thus, the same weights $\alpha$ are assigned to them, while $\sigma_{1221}$ is weighted with $\beta$. 
Depending on the distribution of `$1$'s, the sum for this bipartition is found to be either (a) $\tilde{G}_r^{(1)}=7\alpha$ or (b) $\tilde{G}_r^{(2)}=3\alpha+\beta$. 
The best weights are obtained when the two assignments are equally good, i.e., $7\alpha=3\alpha+\beta$.}
\label{fig:criteria}
\end{figure}
\end{center}

To derive a combined entanglement witness for the GHZ state, we use all eight operators labeled by ${\cal S}$ (see Tab.~\ref{tab:corrsGHZ}). 
We assign equal weights to the seven operators obtained from the measurement setting ${\cal M}_{3333}$, i.e., $\alpha=v_{3333}=v_{0033}=\dots=v_{3300}$ since these mutually commute and behave similarly with regard to the cut-anticommutation relations with $\sigma_{1221}$ for the different bipartitions.
The weight of the remaining operator will be denoted by $\beta=v_{1221}$. 
From the anti-commutativity graph (one without any edges) one obtains $G_0=7\alpha+\beta$.
Depending on the distribution of `$1$'s, the sums for all bipartitions are either $G_r^{(1)}=7\alpha$ or $G_r^{(2)}=3\alpha+\beta$, see Fig.~\ref{fig:criteria}.
For optimal noise robustness, one has to find the weights $v_{\oldvec{j}}$ by minimizing $G/G_0$. 
The minimum is achieved for $G_r^{(1)}=G_r^{(2)}$, thus $7\alpha=3\alpha+\beta$, which leads, by arbitrarily setting $\alpha=1$, to $G_0=7\alpha+\beta=7+4=11$ and $G=7\alpha=3\alpha+\beta=7$. 
Then, the optimized two-measurement-witness for the GHZ state reads
\begin{align}
{\cal W}^{\rm{GHZ}}=\frac{1}{11}\left(T_{3333}^2+T_{3300}^2+T_{0033}^2+T_{3003}^2+T_{0330}^2\right.\nonumber\\
+\left.T_{3030}^2+T_{0303}^2+4T_{1221}^2 \right)\leq_{\rm BISEP}\frac{7}{11}.\label{eq:ghz}
\end{align}
Analogously, for the cluster state $|{\cal C}_4\rangle\propto(|0000\rangle + |0011\rangle - |1100\rangle + |1111\rangle)$ one obtains the witness 
\begin{align}
{\cal W}^{\mathrm{\C}_4}=\frac{1}{6}\left(T_{3300}^2+T_{3011}^2+T_{0311}^2+T_{1130}^2+T_{1103}^2\right.\nonumber\\
+\left.T_{0033}^2 \right)\leq_{\rm BISEP}\frac{2}{3}.
\label{eq:c4}
\end{align}
For details on the derivation, see the SM~\cite{SM}.

\textit{Extensions.---}
Similar criteria can also be formulated for more qubits. 
The two-measurement-witness for the $N$-qubit GHZ state is based upon the measurements of ${\cal M}_{3333\dots3}$ and, e.g., ${\cal M}_{2211\dots1}$ since one is able to find operators whose expectation value can be determined by those measurements such that Eq.~(\ref{eq:complementarity}) can be violated for each bipartition. 
Then, genuine multipartite entanglement is detected by violation of
\begin{align}
{\cal W}^{\mathrm{GHZ}_N}=\frac{1}{2^{N-1}+2^{N-2}-1}\Big[T_{3333\dots3}^2+T_{0033\dots3}^2\nonumber\\
+T_{0303\dots3}^2+\dots+T_{33\dots300}^2+2^{N-2}T_{2211\dots1}^2\Big]\nonumber\\
\leq_{\rm BISEP}\frac{2^{N-1}-1}{2^{N-1}+2^{N-2}-1}\mathop{\longrightarrow}\limits_{N\rightarrow\infty}\frac{2}{3}.
\end{align}
The extension of the criterion for the $N$ qubit cluster state $\ket{\mathrm{\tilde{\C}}_N}$ ($N$ even) is based on the correlations $\{T_j | j\in{\cal S}_{1313\dots13}\cup{\cal S}_{3131\dots31}\}$ where the set ${\cal S}_k$ indexes all nonvanishing correlations of the cluster state that can be determined from the measurement setting ${\cal M}_k$. 
Please note that $\ket{\mathrm{\tilde{\C}}_4}$ as definied via the stabilizer formalism~\cite{Stabilizer} equals $\ket{\mathrm{{\C}}_4}$ up to LU transformations.
Genuine multipartite entanglement of $\ket{\mathrm{\tilde{\C}}_N}$ is then identified by violation of 
\begin{align}
{\cal W}^{\mathrm{\tilde{\C}}_N}=\frac{\sum_{j\in{\cal S}_{1313\dots13}}T_j^2+\sum_{j\in{\cal S}_{3131\dots31}}T_j^2}{2\left(2^{{N}/{2}}-1\right)}\nonumber\\
\leq_{\rm BISEP}\frac{2^{{N}/{2}-1}+2^{{N}/{2}}-2}{2\left(2^{{N}/{2}}-1\right)}\mathop{\longrightarrow}\limits_{N\rightarrow\infty}\frac{3}{4}.
\end{align}

%%%%%% EXPERIMENTAL PART %%%%%%%

\textit{Analysis of experimental data.}---In order to experimentally demonstrate the applicability of our new entanglement criteria, we prepare a series of superpositions of GHZ and cluster states with variable weights. 
Different linear optical setups to prepare either four-qubit GHZ~\cite{GHZprepare} or cluster states~\cite{Cluster} are known.
To have the flexibility to prepare superpositions of GHZ and cluster states in a single setup, we resort to a two photon experiment using two degrees of freedom per photon, namely polarization and path. 
This approach enables one to prepare states with both high fidelity and high count rates.
From now on, the computational basis $|0\rangle$ and $|1\rangle$ is encoded either in the polarization or in the path degree of freedom, i.e. $|0\rangle \longrightarrow |H\rangle$ and $|1\rangle \longrightarrow |V\rangle$ for horizontal ($H$) and vertical ($V$) polarization and  $|0\rangle \longrightarrow |a\rangle$ and $|1\rangle \longrightarrow |b\rangle$ for paths $a$ and $b$.\\
\begin{figure}[!t]
\includegraphics[width=0.40\textwidth]{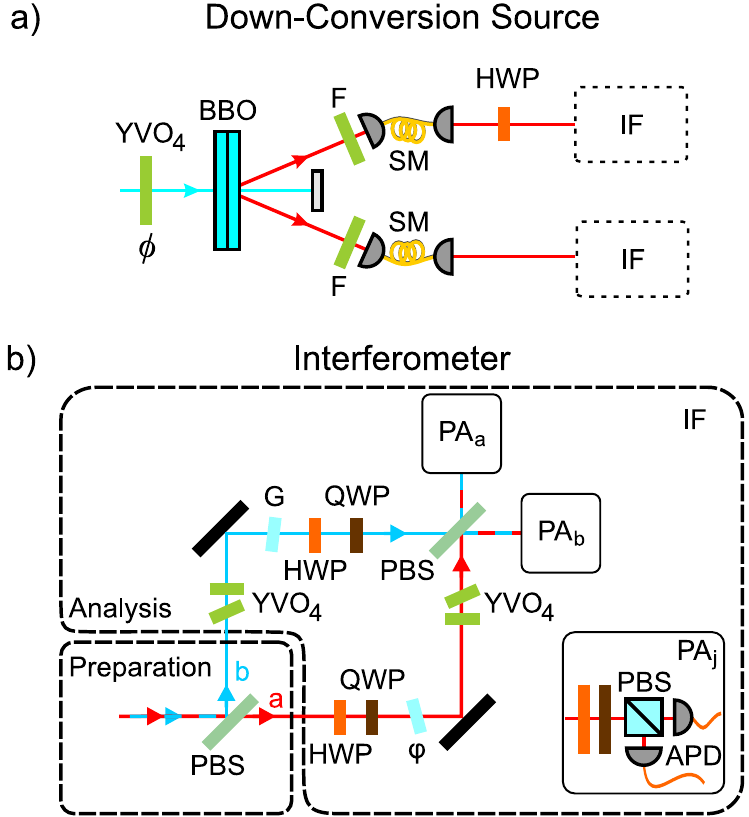}
\caption{Scheme of the experimental setup. 
In a first step (a) a type-I SPDC source together with a half waveplate (HWP) at angle $\theta$ is used to prepare states of the form 
$\bigl(|H\rangle(\cos2\theta |H\rangle + \sin2\theta |V\rangle) + {\rm e}^{i\phi}|V\rangle(-\cos2\theta |V\rangle + \sin2\theta |H\rangle)\bigr)/{\sqrt{2}}$. 
The phase $\phi$ can be set by a birefringent yttrium-vanadate crystal (YVO$_4$). Interference filters (F) are applied for spectral filtering and spatial filtering is performed by coupling into single mode fibers (SM~\cite{SM}).
In a second step (b), the state preparation is completed by increasing the Hilbert space by polarizing beam splitters (PBS). 
Overlap at a beamsplitter and polarization analysis allows to measure all Pauli settings $\sigma_i$ and to perform QST.
YVO$_4$ crystals and glass plates (G and $\varphi$) inside the interferometer are used for phase and path length compensation, respectively.
}
\label{fig:source}
\end{figure}
The photon source shown in Fig.~\ref{fig:source}(a) uses spontaneous parametric down-conversion and allows to prepare states of the form $\bigl(|H\rangle(\cos(2\theta) |H\rangle + \sin(2\theta) |V\rangle) + {\rm e}^{i\phi}|V\rangle(\sin(2\theta) |H\rangle - \cos(2\theta) |V\rangle)\bigr)/{\sqrt{2}}$ (see the SM~\cite{SM} for details).
In order to achieve the intended four-qubit state, coupling to the path degree of freedom is required. 
Thus, the polarization dependence of the output of a polarizing beamsplitter is used, i.e., photons are transformed as $|H\rangle \longrightarrow |H a\rangle$ and $|V\rangle \longrightarrow |V b\rangle$ with $a$ and $b$ denoting the corresponding output modes of the PBS, see Fig.~\ref{fig:source}(b).
Consequently, four-qubit states parametrized by $\theta$ and $\phi$, $|\Psi\left(\theta,\phi\right)\rangle=\big(\cos(2\theta) |HaHa\rangle + \sin(2\theta) |HaVb\rangle + \operatorname{e}^{i\phi}\sin(2\theta) |VbHa\rangle - \operatorname{e}^{i\phi}\cos(2\theta) |VbVb\rangle\big)/{\sqrt{2}}$, is obtained. 
Prominent members of $|\Psi\left(\theta,\phi\right)\rangle$ are for example the GHZ states $(|HaHa\rangle \mp |VbVb\rangle)/{\sqrt{2}}$ for $\theta=0$ and $\phi = 0,\pi$, respectively, or the cluster states $(|HaHa\rangle + |HaVb\rangle \pm |VbHa\rangle \mp |VbVb\rangle)/{{2}}$ obtained for $\theta=\pi/8$ and $\phi = 0,\pi$.

The prepared states are characterized by means of QST, proving full control of the experimental apparatus.
This can be achieved with an interferometer setup as shown in Fig.~\ref{fig:source}(b), overlapping the modes $a$ and $b$ together with a polarization analysis and coincidence detection in the outputs.

\textit{Experimental results.}---$13$ states were prepared with $\phi=\pi$ and $\theta$ being increased from $0$ (GHZ) to ${\pi}/{8}$ (cluster) and to ${\pi}/{4}$ (GHZ$^\prime$) in equidistant steps.
The coincidence rate was approximately $100\,{\rm s}^{-1}$ with a measurement time of $40\,{\rm s}$ for each basis setting, resulting in 3700$-$4400 counts per setting and a measurement time of about $12\,{\rm h}$ to perform QST for all states. 
A measure for the quality of a prepared state $\varrho_{\rm exp}$ with respect to a pure target state $|\psi\rangle$ is the fidelity ${\cal F} = \Tr(\varrho_{\rm exp} |\psi\rangle\langle\psi|)$. 
For the GHZ state, we observed a fidelity of ${\cal F} = 0.958 \pm 0.004$, while for the cluster state it was ${\cal F} = 0.962 \pm 0.003$.
For the other states, see Tab.~\ref{tab:statesResults} in the SM~\cite{SM}.

\begin{figure}[!t]
\includegraphics[width=0.40\textwidth]{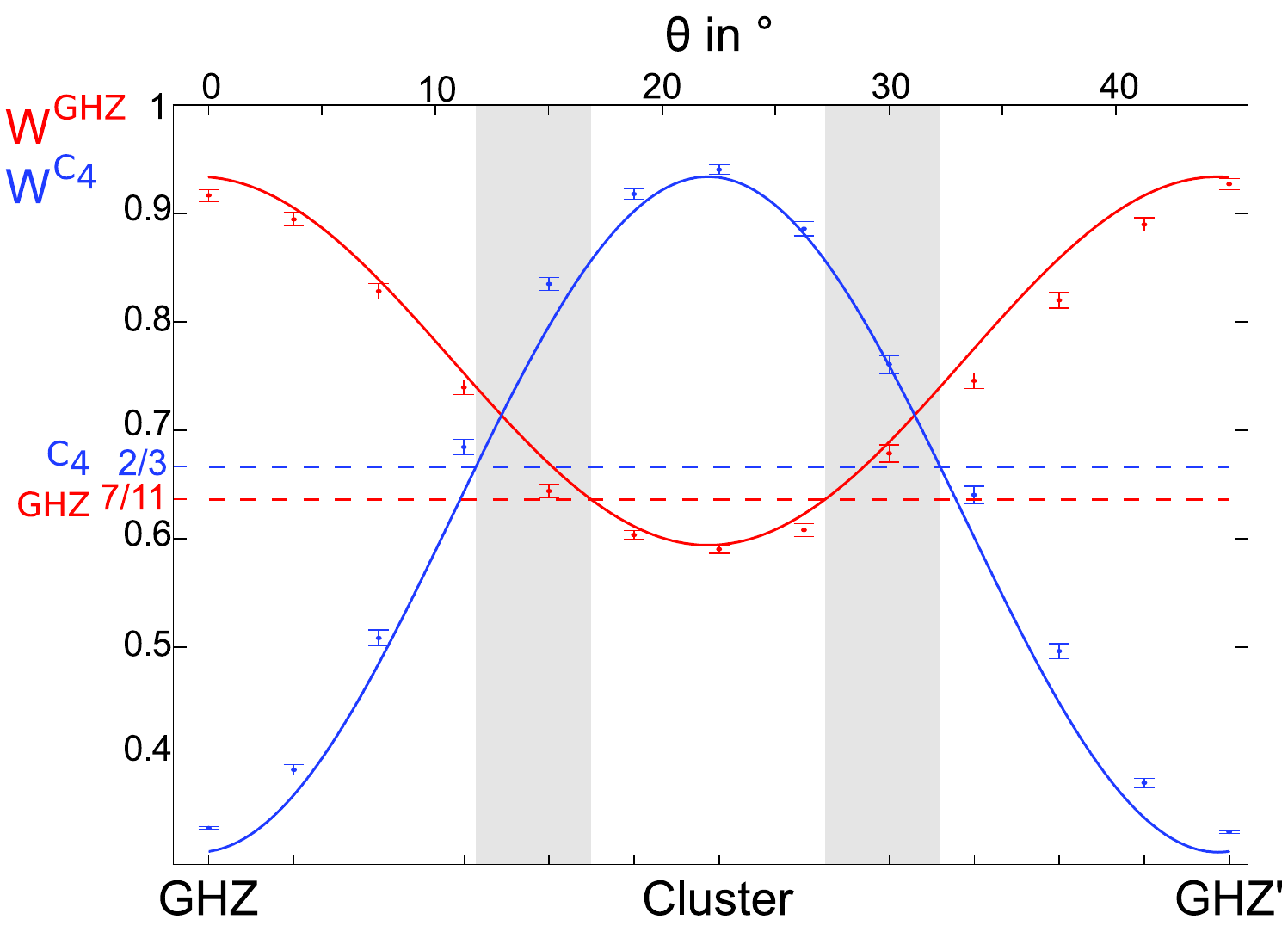}
\caption{The entanglement criterion for the GHZ state allows to detect most of the superpositions of GHZ and cluster state to be genuinely four-partite entangled (red) whereas the criterion for the cluster state detects states around $\theta=22.5^{\circ}$ to be genuinely four-partite entangled (blue). 
States within the gray shaded areas can be detected to be genuinely four-partite entangled by means of both criteria. 
The solid lines show the theoretically expected values for the target states $|\Psi\left(\theta,\phi\right)\rangle$ mixed with white noise such that on average the fidelities correspond to the measured values.
}
\label{fig:criterion}
\end{figure}

Genuine four-partite entanglement could be tested using two measurement settings only. 
Let us start to determine the witnesses for the GHZ state from measuring two settings ${\cal M}_{3333}$ and ${\cal M}_{1221}$.
The values of the respective measured correlations (Tab.~\ref{tab:expCorr} in the SM~\cite{SM})
lead to a violation of all seven criteria by at least $56$ standard deviations for all cuts, see Tab.~\ref{tab:expResults}. 
\begin{table}[b]
\centering
\caption{Experimental values of the individual criteria and combined witnesses for the considered states. 
All values for a specific bipartition are clearly above the threshold of $1/2$, indicating genuine four-partite entanglement in all cases.
The thresholds for the combined criteria are $7/11$ (GHZ), $2/3$ (cluster), $4/5$ ($D_4^{(2)}$), and $3/5$ ($\Psi_4$) respectively.
The Dicke state cannot be significantly proven to be genuinely fourpartite entangled by means of the combined witness, see SM~\cite{SM}.
Hence, one has to resort to the individual criteria in this case.
}
\vspace{0.2cm}
\resizebox{\columnwidth}{!}{%
\begin{tabular}[b]{|l |c |c| c|c |} 
\hline \hline 
Partition	& $\ket{\rm GHZ}$ & $\ket{\C_4}$ & $\ket{D_4^{(2)}}$ & $\ket{\Psi_4}$ \\
\hline 
$A|BCD$		&$0.894\pm0.007$&$0.922\pm0.006$&$0.819\pm0.013$&$0.804\pm0.019$ \\
$B|ACD$		&$0.906\pm0.006$&$0.940\pm0.004$&$0.819\pm0.013$&$0.804\pm0.019$ \\
$C|ABD$		&$0.906\pm0.006$&$0.940\pm0.004$&$0.819\pm0.013$&$0.804\pm0.019$ \\
$D|ABC$		&$0.906\pm0.006$&$0.928\pm0.006$&$0.819\pm0.013$&$0.804\pm0.019$ \\
$AB|CD$		&$0.904\pm0.006$&$0.922\pm0.006$&$0.627\pm0.013$&$0.608\pm0.017$ \\
$AC|BD$		&$0.906\pm0.006$&$0.948\pm0.004$&$0.620\pm0.013$&$0.594\pm0.021$ \\
$AD|BC$		&$0.901\pm0.006$&$0.943\pm0.004$&$0.625\pm0.013$&$0.622\pm0.021$\\ \hline\hline
combined	&$0.916\pm0.005$&$0.940\pm0.004$&$0.801\pm0.017$&$0.683\pm0.014$\\
\hline \hline
\end{tabular}%
\label{tab:expResults}%
}
\end{table}
Also, the combined criterion ${\cal W}^{\mathrm{GHZ}} = 0.916 \pm 0.005 > \frac{7}{11}$ certifies genuine four-partite entanglement. 
For the cluster state, according to our entanglement criterion, the measurement settings ${\cal M}_{1133}$ and ${\cal M}_{3311}$ were used (see SM~\cite{SM}), resulting in
${\cal W}^{\mathrm{\C}_4} = 0.940 \pm 0.004 >\frac{2}{3}$ for the combined criterion.

Using the combined witnesses, we analyze the entanglement for all states $|\Psi\left(\theta,\phi\right)\rangle$ (Fig.~\ref{fig:criterion}).
As can be seen, $10$ of $13$ states can be detected as genuinely four-partite entangled by the criterion ${\cal W}^{\rm GHZ}$, the $6$ states close to the cluster state can be determined by means of ${\cal W}^{\mathrm{\C}_4}$. 
Some states can be shown to be truly four-partite entangled by means of both criteria as both are above their respective threshold.
Genuine fourpartite entanglement could be proven with experimental data of the Dicke state $|D_4^{(2)}\rangle$~\cite{NoCorrs} and the singlet state~\cite{Familypaper}, see Tab.~\ref{tab:expResults}. 
For more details see the SM~\cite{SM}.

\textit{Conclusion.---}We have introduced a novel scheme for the systematic construction of entanglement witnesses, which need a minimal number of measurements for their evaluation independent of the number of qubits. 
We believe that such a minimal multipartite entanglement detection will become a handy diagnostic procedure as it is fast and simple. 
It is an interesting question what other states can reveal their multipartite quantum correlations in two measurements. 
Another challenge is to find even stronger criteria, which, by possibly going to few more measurements, will detect multipartite entanglement with a higher robustness against noise.

\textit{Acknowledgments.---}The work is subsidized form funds for science for years 2012-2015 
approved for
international co-financed project BRISQ2 by Polish Ministry of Science 
and Higher Education (MNiSW).
We thank D. Richart and S. Habib for stimulating discussions.
MW was supported by FNP project HOMING-PLUS/2011-4/14 and later by MNiSW Grant IdP2011 00361 (Ideas Plus). 
We acknowledge the support of this work by the EU (QWAD, ERC QOLAPS, FP7 BRISQ2). 
CS and LK thank QCCC and ExQM, respectively, of the Elite Network of Bavaria for support.

\newpage

\section*{\large Supplemental Material}

\twocolumngrid
\section*{SM\,1: Constructing optimal criteria}
Criteria to detect genuine n-partite entanglement are specifically designed for individual states. 
Here, we will describe the construction of the criteria for the cluster state $|{\cal C}_4\rangle\propto(|0000\rangle + |0011\rangle - |1100\rangle + |1111\rangle)$. 
Furthermore, we will derive criteria for the symmetric four-qubit Dicke state $|D_4^{(2)}\rangle\propto |1100\rangle+|1010\rangle+|1001\rangle+|0110\rangle+|0101\rangle+|0110\rangle$, the four-qubit singlet state $|\Psi_4\rangle\propto |0011\rangle+|1100\rangle-1/2\left(|0110\rangle+|1001\rangle+|0101\rangle+|1010\rangle\right)$ and for $|W_4\rangle\propto |1000\rangle+|0100\rangle+|0010\rangle+|0001\rangle$.
\subsection{Cluster state}
\begin{center}
\begin{table}[b]
\caption{The correlations of the cluster state. 
All other correlations vanish in the Pauli basis. 
Those with blue characters can be inferred from the setting ${\cal M}_{1133}$, the red colored ones from ${\cal M}_{3311}$.}
\vspace{0.3cm}
\begin{tabular}[b]{|c |r |c| c|r |c|c|r|c|c|r|} 
%\multicolumn{4}{c}{Cluster state} \\
\hline \hline 
$T_{0000}$ 			& $1$  &\textcolor{blue}{$\boldsymbol{T_{0033}}$} 	& $1$ &
$\textcolor{red}{\boldsymbol{T_{0311}}}$	& $1$  &$T_{0322}$ 			& $-1$ \\
$\textcolor{blue}{\boldsymbol{T_{1103}}}$ 	& $-1$ &$\textcolor{blue}{\boldsymbol{T_{1130}} }$	& $-1$ &
$T_{1212}$ 			& $-1$ &$T_{1221}$ 			& $-1$ \\
$T_{2112}$ 			& $-1$ &$T_{2121}$ 			& $-1$ &
$T_{2203}$			& $1$ &$T_{2230}$ 			& $1$ \\
$\textcolor{red}{\boldsymbol{T_{3011}} }$	& $1$ &$T_{3022}$ 			& $-1$ &
$\textcolor{red}{\boldsymbol{T_{3300}} }$ 	& $1$ &$T_{3333}$ 			& $1$ \\
\hline \hline
\end{tabular}
\label{tab:corrsCluster}
\end{table}
\end{center}
The general procedure of finding the optimal entanglement criteria has already been described in the main text, together with an illustrative example for the GHZ state. 
According to the scheme given in the main text, Tab.~\ref{tab:corrsCluster} lists the non-vanishing correlations of the cluster state.
%One notices that measurements of the settings $xxzz$ and $zzxx$ are sufficient to infer six of the non-zero correlations of the state.
One notices that the settings ${\cal M}_{1133}$ and ${\cal M}_{3311}$ are sufficient to infer six of the non-zero correlations of the state, indicated by bold letters in Tab.~\ref{tab:corrsCluster}.
We use the corresponding operators to build the set $\{\sigma_{1103}$, $\sigma_{1130}$, $\sigma_{0033}$, $\sigma_{0311}$,  $\sigma_{3011}$, $\sigma_{3300}\}$.
% \begin{align}
% {\cal S}_{{\cal C}_4} &=&\{&\sigma_0\sigma_3\sigma_1\sigma_1,\sigma_3\sigma_0\sigma_1\sigma_1,\sigma_3\sigma_3\sigma_0\sigma_0,\nonumber\\
% &&&\sigma_1\sigma_1\sigma_0\sigma_3,\sigma_1\sigma_1\sigma_3\sigma_0,\sigma_0\sigma_0\sigma_3\sigma_3\}.
% \end{align}
%For simplicity, we use ordinal numbers for labeling the operators of ${\cal S}_{{\cal C}_4}$, i.e., $o_1=\sigma_0\sigma_3\sigma_1\sigma_1$, $o_2=\sigma_3\sigma_0\sigma_1\sigma_1$, $\dots$, $o_6=\sigma_0\sigma_0\sigma_3\sigma_3$. 
Indeed, all six operators mutually commute and we are able to find cut-anticommutation relations for each bipartition.
Therefore, it is possible  to disprove separability along each cut by using an inequality like Eq.~(\ref{eq:complementarity}). 
For example, the first two operators $A|BCD$-anticommute with the last two, so $A|BCD$-nonseparability is proven by violation of
\begin{equation}
{\cal W}^{C_4}_{A|BCD}=\left(T_{1103}^2+ T_{3011}^2\right)\leq_{\mathop{\rm SEP}\limits_{A|BCD}} 1.
\label{eq:ClusterA_BCD1}
\end{equation}
or any other combination of squared expectation values of the operators of these two groups. 
Averaging over all possible combinations of $A|BCD$-anticommuting operators of the given set and normalization leads to
\begin{align}
{\cal W}^{C_4}_{A|BCD}=\frac{1}{2}\Big[&{\textstyle \frac{1}{2}}\left(T_{1103}^2+T_{1130}^2\right)\nonumber\\
+&{\textstyle \frac{1}{2}} \left(T_{3011}^2+T_{3300}^2\right)\Big]\leq_{\mathop{\rm SEP}\limits_{A|BCD}}\frac{1}{2}.
\label{eq:ClusterA_BCD_Av}
\end{align}
Permutations of the indices used in Ineq.~(\ref{eq:ClusterA_BCD_Av}) are used for other 1:3 cuts and for $AB|CD$.
% Note that Ineq.~(\ref{eq:ClusterA_BCD}) can also be used to verify non-separability along all 2:2 cuts. 
For cuts ${\cal B}^{*}=\{AC|BD,AD|BC\}$, the separability is refuted more efficiently with
% Separability along ${\cal B}^{*}=\{AC|BD,AD|BC\}$ is refuted with
\begin{align}
{\cal W}^{C_4}_{{\cal B}^{*}}=\frac{1}{6}\Big[&T_{0033}^2+T_{0311}^2+T_{1103}^2\nonumber\\
+&T_{1130}^2+T_{3011}^2+T_{3300}^2\Big]\leq_{\mathop{\rm SEP}\limits_{{\cal B}^{*}}}\frac{1}{2}.
\label{eq:ClusterAC_BD}
\end{align}

\begin{figure}
{\includegraphics[width=0.47\textwidth]{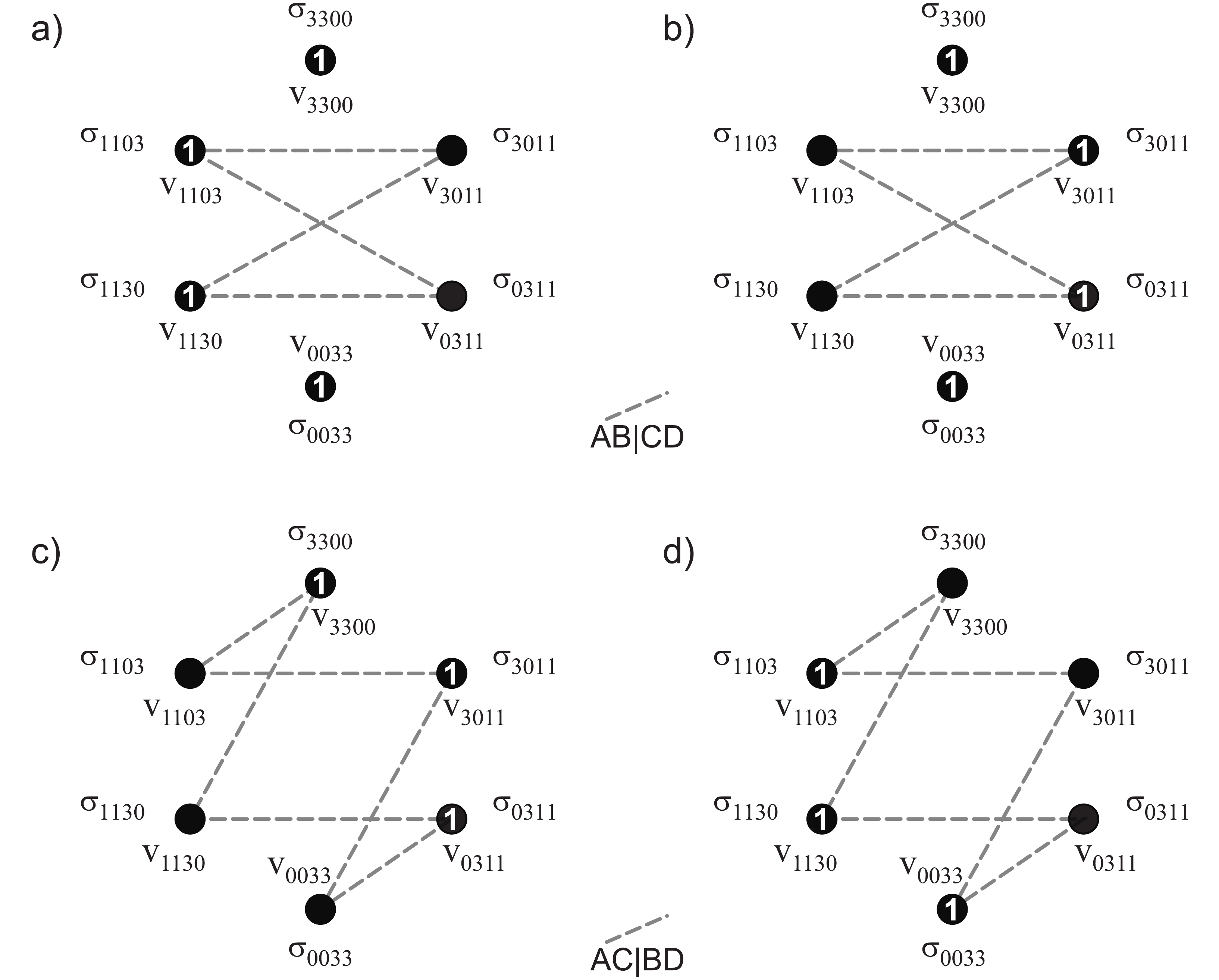}}
\caption{The cut-anticommutativity graphs for the cluster state are shown for $AB|CD$ (upper row) and $AC|BD$ (lower row). 
For those bipartitions, for which the cut-anticommutation relations are indicated by dashed lines, two possible distributions of values `$1$'s are depicted in the respective columns.
The weights of the operators are denoted by, e.g., $v_{1103}$.}
\label{fig:criteria_drawing_cluster}
\end{figure}

The combined entanglement witness for the cluster state uses all six operators. 
Therefore, the anticommutativity graphs consist of six vertices, each representing one of the operators of the given set.
For no bipartition ($r=0$), there are no edges and  `$1$'s can be assigned to all vertices. 
Thus, this graph leads to the sum $G_0=v_{0311}+v_{3011}+v_{3300}+v_{1103}+v_{1130}+v_{0033}$. 
The graph of the bipartition $AB|CD$ is depicted in Fig.~\ref{fig:criteria_drawing_cluster} a) and b), where the dashed lines indicate the cut-anticommutation relations.
Because neither $\sigma_{0033}$ nor $\sigma_{3300}$ $AB|CD$-anticommutes with any of the operators, they can be assigned `$1$' in any case.
Besides this, one can distribute `$1$'s according to Fig.~\ref{fig:criteria_drawing_cluster} a) leading to ${G}_{AB|CD}^{(1)}=v_{3300}+v_{1103}+v_{1130}+v_{0033}$. 
A different distribution of `$1$'s results in ${G}_{AB|CD}^{(2)}=v_{0311}+v_{3011}+v_{3300}+v_{0033}$, see Fig.~\ref{fig:criteria_drawing_cluster} b). 
The operators $\sigma_{3300}$ and $\sigma_{0033}$, appearing in both distributions in this cut, seem to be superior [not connected with any operator in Fig.~\ref{fig:criteria_drawing_cluster} a) and b)] to the other operators. 
In contrast, e.g., for the bipartition $A|BCD$, the operators $\sigma_{0311}$ and $\sigma_{0033}$ do not cut-anticommute with any other operators and are, thus, superior for this bipartition. 
Considering the bipartition $AC|BD$, as it is shown in Fig.~\ref{fig:criteria_drawing_cluster} c) and d), the cut-anticommutativity relations are such that each operator cut-anticommutes with two operators. 
Thus, by considering all seven bipartitions, all six operators behave similarly, suggesting equal weights for all operators, i.e. we introduce $\alpha$ with $\alpha=v_{0311}=v_{3011}=\dots=v_{0033}$. 
Without loss of generality, we can set $\alpha=1$. 
Because four `$1$'s can be distributed for the bipartition $AB|CD$, our chosen weights result in ${G}_{AB|CD}^{(1)}={G}_{AB|CD}^{(2)}=4$.
In conclusion, we find for the anticommutativity graph $G_0=6\alpha=6$ and by maximizing over all bipartitions ${G}=4\alpha=4$, leading to the criterion Eq.~(\ref{eq:c4}).
% \begin{align}
% I_{\mathrm{\cal C}_4}=\frac{1}{6}\left(T_{0311}^2\right.+T_{3011}^2&+&T_{3300}^2+T_{1103}^2+T_{1130}^2\nonumber\\
% &+&\left.T_{0033}^2 \right)\leq_{\rm BISEP}\frac{2}{3}.\label{eq:c4}
% \end{align}

\subsection{Dicke state}
% Graph states like the GHZ or cluster states, where all non-vanishing correlations are $\pm 1$, can easily be detected to be genuinely multipartite entangled with our scheme. 
% In contrast, some states have correlations too weak to reveal their nonclassical features in our approach when all bipartitions are covered by a single criterion.
% Nevertheless, we can modify our approach to detect multipartite entanglement in states with weaker correlations.
% The idea is to construct individual indicators specifically for each cut such that all witnesses can be computed from the same experimental data. 
% If all these criteria for non-separability will be violated at the same time, we can prove genuine multipartite entanglement for that state. 
Criteria for the symmetric four-qubit Dicke state with two excitations $|D_{4}^{(2)}\rangle\propto |1100\rangle+|1010\rangle+|1001\rangle+|0110\rangle+|0101\rangle+|0110\rangle$ can also be derived easily. 
%Nevertheless, caution has to be practiced, since only $T_{1111}^2$, $T_{2222}^2$, and $T_{3333}^2$ are $1$ for $\ket{D_4^{(2}}$ while all other correlations are less. 
Measurement of ${\cal M}_{1111}$ enables to deduce the expectation values of the operators $\sigma_{0011}$, $\sigma_{0101}$, $\sigma_{0110}$, $\sigma_{1001}$, $\sigma_{1010}$, $\sigma_{1100}$, and $\sigma_{1111}$. 
Adding the measurement of ${\cal M}_{2222}$ results in a set of commuting operators also containing cut-anticommuting operators for every cut. 
The possibility of one-versus-three qubit separability for the Dicke state is eliminated by a violation of
\begin{equation}
{\cal W}^{D_4^{(2)}}_{\{1:3\}}=\frac{1}{2}\left(T_{1111}^2+T_{2222}^2\right)\leq_{\mathop{\rm SEP}\limits_{\{1:3\}}}\frac{1}{2}.
\label{eq:dicke1}
\end{equation}
Since only $T_{1111}^2$ and $T_{2222}^2$ are $1$ for $\ket{D_4^{(2}}$ while all further expectation values of operators deduced by those two are less, one cannot enhance the criterion by averaging over additional values as in, e.g., Eq.~(\ref{eq:GHZ_A_BCD}). 
To rule out separability along cut, say, $AB|CD$ the inequality 
\begin{align}
{\cal W}^{D_4^{(2)}}_{AB|CD}=\frac{1}{2}\Big[{\textstyle\frac{1}{4}}\left(T_{1010}^2+T_{1001}^2+T_{0110}^2+T_{0101}^2\right)\nonumber\\
+T_{2222}^2\Big]\leq_{\mathop{\rm SEP}\limits_{AB|CD}}\frac{1}{2}
\label{eq:dicke2}
\end{align}
has to be violated. 
Note that even the ideal Dicke state  will score only ${\cal W}^{D_4^{(2)}}_{AB|CD}=13/18\approx 0.72$.
By permutations of the indices in Eq.~(\ref{eq:dicke2}) one obtains the criteria ${\cal W}^{D_4^{(2)}}_{AC|BD}$ and ${\cal W}^{D_4^{(2)}}_{AD|BC}$.
The combined witness for the Dicke state reads
\begin{align}
{\cal W}^{D_4^{(2)}}=\frac{1}{10}\Big[2\left(T_{1111}^2+T_{2222}^2\right)\nonumber\\
+T_{1100}^2+T_{0011}^2+T_{0101}^2+T_{1010}^2+T_{1001}^2+T_{0110}^2\nonumber\\
+T_{2200}^2+T_{0022}^2+T_{0202}^2+T_{2020}^2+T_{2002}^2+T_{0220}^2\Big]\leq_{\mathop{\rm SEP}}\frac{4}{5},
\label{eq:dickeCombined}
\end{align}
which is difficult to violate experimentally, as even the ideal Dicke state scores only $14/15$. % as the ideal Dicke state $\ket{D_4^{(2)}}$.
Since our data cannot show a significant proof of genuine fourpartite entanglement, all possible biseparations have to be ruled out.
% Though it would be possible to construct a single combined witness also for the Dicke state, it would be very difficult to violate experimentally; even the ideal Dicke state will score poorly because of it's relatively small correlations.

% For example, the criterion 
% \begin{equation}
% %I_{D_4^{(2)},1}=
% I_{D1}=\frac{1}{2}\left(T_{1111}^2+T_{3333}^2\right)\leq_{\mathop{\rm SEP}\limits_{\{1:3\}}}\frac{1}{2}
% \label{eq:dicke1}
% \end{equation}
% detects non-separability against all one-versus-three cuts, e.g., $A|BCD$.
% On the other hand, entanglement along the bipartition $AB|CD$ cannot be detected by Eq.~(\ref{eq:dicke1}). 
% Instead, for $AB|CD$ one can deploy the criterion of Eq.~(\ref{eq:dicke2}) 
% % \begin{align}
% % I_{D2}=&\frac{1}{8}&\left(T_{1010}^2+T_{1001}^2+T_{0110}^2+T_{0101}^2\right.\nonumber\\
% % &+&\left.4T_{3333}^2\right)\leq_{\mathop{\rm SEP}\limits_{AB|CD}}\frac{1}{2},
% % \label{eq:dicke2}
% % \end{align}
% which can be evaluated with the same measurement data as Eq.~(\ref{eq:dicke1}), namely $\sigma_{1111}$ and $\sigma_{3333}$. %namely measurements performed in settings $xxxx$ and $zzzz$. 
% Similarly, for the other bipartitions $AC|BD$ and $AD|BC$ criteria ${\cal W}^{D3}$ and ${\cal W}^{D4}$, respectively, are found by permutations of ${\cal W}^{D2}$.

\subsection{Singlet state}
For the four-qubit singlet state $|\Psi_4\rangle \propto |0011\rangle+|1100\rangle-1/2\left(|0110\rangle+|1001\rangle+|0101\rangle+|1010\rangle\right)$, the set of correlations for the operators deduced from the measurement settings of ${\cal M}_{1111}$ and ${\cal M}_{2222}$ are similar as for $|D_{4}^{(2)}\rangle$. 
Thus, the criteria ${\cal W}^{D_4^{(2)}}_{\{1:3\}}$ and ${\cal W}^{D_4^{(2)}}_{AB|CD}$ also apply here while the criteria for the bipartitions $AC|BD$ and $AD|BC$ are slightly modified since $T_{0011}$ and $T_{1100}$ reach a value of only $1/3$ for the singlet state and are therefore left out.
% 
% For the four-qubit singlet state $|\Psi_4\rangle$, one-versus-three-separability can also be ruled out by the criterion ${\cal W}^{\Psi}_{\{1:3\}}={\cal W}^{D_4^{(2)}}_{\{1:3\}}$. 
% The other three possiblities of separability can be eliminated by the criteria
% \begin{align}
% {\cal W}^{\Psi2}&=&\frac{1}{2}\bigg[T_{3333}^2+\frac{1}{4}\left(T_{0101}^2+T_{0110}^2\right.&\nonumber\\
% &&+T_{1001}^2+\left.T_{1010}^2\right)\bigg]&\leq_{\mathop{\rm SEP}\limits_{AB|CD}}\frac{1}{2},\nonumber\\
% {\cal W}^{\Psi3}&=&\frac{1}{2}\left[T_{3333}^2+\frac{1}{2}\left(T_{0110}^2+T_{1001}^2\right)\right]&\leq_{\mathop{\rm SEP}\limits_{AC|BD}}\frac{1}{2},\nonumber\\
% {\cal W}^{\Psi4}&=&\frac{1}{2}\left[T_{3333}^2+\frac{1}{2}\left(T_{1010}^2+T_{0101}^2\right)\right]&\leq_{\mathop{\rm SEP}\limits_{AD|BC}}\frac{1}{2}.
% \end{align}
% A violation of all four criteria detects genuine four-partite entanglement.
% The evaluation of the criteria for $|D_4^{(2)}\rangle$ and $|\Psi_4\rangle$ with experimental data is described in the main text. 
% Indeed, both states can be detected to be truly genuine four-partite entangled with only two measurement settings.
The combined witness for the singlet state reads 
\begin{align}
{\cal W}^{\Psi_4}=\frac{1}{10}\Big[4T_{1111}^2+2T_{3333}^2\nonumber\\
+T_{3003}^2+T_{0330}^2+T_{3030}^2+T_{0303}^2\Big]\leq_{\mathop{\rm SEP}}\frac{3}{5}.
\label{eq:singletCombined}
\end{align}
The ideal state $\ket{\Psi_4}$ scores $7/9$.
The experimentally prepared state could be proven to be genuinely fourpartite entangled with high significance.
\subsection{W state}
Because correlations of $|W_4\rangle\propto|0001\rangle+|0010\rangle+|0100\rangle+|1000\rangle$ are (besides of $T_{3333}=-1$) at most only $\pm1/2$ and thus too weak for a robust combined criterion, we again have to find criteria for the different bipartitions in order to build sensitive indicators.
Entanglement along the cut $A|BCD$ can be detected by violation of
\begin{align}
{\cal W}^{W_4}_{A|BCD}&=\frac{1}{2} \bigg[T_{3333}^2 \nonumber \\
&+ \frac{1}{3}\left(T_{1001}^2+T_{1010}^2+ T_{1100}^2 \right)\bigg]\leq_{\mathop{\rm SEP}\limits_{A|BCD}}\frac{1}{2}.
\end{align}
Criteria for the other one-versus-three-separations are obtained by permuting the parties in ${\cal W}^{W_4}_{A|BCD}$. 
$AB|CD$-separability can be ruled out by the criterion 
\begin{align}
{\cal W}^{W_4}_{AB|CD}&=&\frac{1}{2} \bigg[ T_{3333}^2 + \frac{1}{4}\left(T_{0101}^2 +T_{0110}^2\right. &\nonumber\\
&&\left. + T_{1001}^2+T_{1010}\right)\bigg]&\leq_{\mathop{\rm SEP}\limits_{AB|CD}}\frac{1}{2},
\end{align}
whose permutations lead to the criteria to eliminate separability along $AC|BD$ and $AD|BC$.

% \subsection{Improvements for GHZ and cluster state}
% Considering each biseparation individually improves the entanglement indicators also for GHZ and $|C_4\rangle$ states.
% %so that the threshold for visibility is $v_{crit}=\sqrt{1/2}\approx 0.707$. 
% For example
% \begin{equation}
% %I_{D_4^{(2)},1}=
% {\cal W}^{\mathrm{GHZ}1}=\frac{1}{2}\left(T_{1221}^2+T_{3333}^2\right)\leq_{\mathop{\rm SEP}\limits_{\{1:3\}}}\frac{1}{2}
% \label{eq:criterionGHZ1}
% \end{equation}
% covers all cuts with a single qubit on one side. 
% Because of the threshold being $1/2$, a visibility of $\sqrt{1/2}\approx0.707$ for the correlations $T_{1221}$ and $T_{3333}$ is sufficient to rule out one-versus-three-separability. 
% For each criterion, a set of operators can be found such that two subsets of mutually commuting operators exist where the operators of the first subset cut-anticommute with those of the second subset. 
% With properly assigned weights, the criterion can obtain a threshold of $1/2$. 
% Naturally, this improvement would come at the price of a larger effort in terms of processing the data. 
% But since each of these indicators for individual cuts involves only two operators, the difficulty of analyzing the anticommutativity graph is removed.
\FloatBarrier
\twocolumngrid
\section*{SM\,2: Setup and Measurement}
The general idea of the experimental setup was already explained in the main text of this letter. Here, we want to 
focus on the details of both the spontaneous parametric down conversion (SPDC) source and the interferometers 
which allow to perform a complete tomographic analysis of the prepared states.

The experiment starts with the generation of pairs of polarization entangled photons in the state 
\begin{equation}
 \frac{1}{{\sqrt{2}}}\left(|HH\rangle + {\rm e}^{i\phi}|VV\rangle\right)
\end{equation}
as obtained from the process of spontaneous parametric down conversion (SPDC)~\cite{PavelJames}.
The SPDC source consists of a pair of crossed type I cut $\beta$-Barium-Borate (BBO) crystals that are pumped by a continuous wave laser diode at a central wavelength of 402\,nm, with approximately $60\,$mW of pump power, and linear polarization of 45$^{\circ}$. 
The phase $\phi$ between the emitted photons can be set by means of an Yttrium Vanadate crystal (YVO$_{4}$) in front of the BBO crystals, see Fig. 2 of the main text. 
An additional half waveplate set at an angle $\theta$ enables to rotate the polarization of the second photon to any linear polarization, leading to the state 
\begin{align}
 \frac{1}{\sqrt{2}} \big(&|H\rangle(\cos(2\theta) |H\rangle + \sin(2\theta) |V\rangle) \notag\\ 
 + {\rm e}^{i\phi}&|V\rangle(\sin(2\theta) |H\rangle - \cos(2\theta) |V\rangle)\big)
\end{align}
as mentioned in the main text.

The emitted photons are spectrally filtered by  interference filters with a bandwidth of 5\,nm. 
Spatial filtering is achieved by coupling the pairs into two single mode fibers that are connected to one of the input ports of each of the two interferometers. 

In principle, a Mach-Zehnder type interferometer as given in the main text suits the purpose to analyze the phase 
between the two spatial modes $a$ and $b$. However, in terms of phase stability, a Sagnac configuration is preferable 
to a Mach-Zehnder type interferometer if one wants to avoid using an active stabilization scheme, as is the case here. 
\begin{figure}[!ht]
{\includegraphics[width=0.45\textwidth]{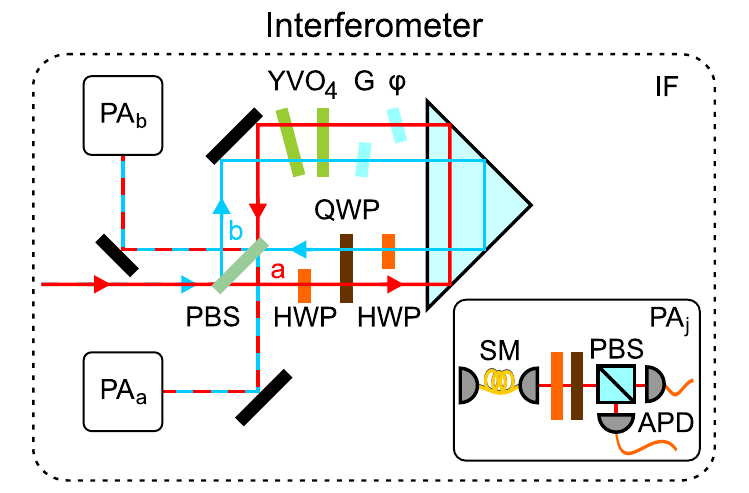}}
\caption{For state analysis, it is necessary to be able to characterize the phase $\varphi$ between the modes 
$a$ and $b$. In principle, a Mach-Zehnder interferometer can be used to measure the (relative) phase between the 
modes $a$ and $b$, but in terms of stability a Sagnac configuration is preferable. Therefore, for the experimental 
realization of the interferometer, we resorted to a Sagnac loop, where instead of two polarizing beam splitters 
(PBS) a single one that is hit by the photons twice is sufficient. Half (HWP) and quarter waveplates (QWP) are 
used for state analysis and enable tomographic analysis of the experimentally prepared state. A pair of YVO$_4$ 
crystals is used for the compensation of unwanted additional phase shifts resulting from the total internal 
reflection at the prism. Two thin glass plates (G) of which one is motorized is used to balance the interferometer 
arms and to set arbitrary relative phases  $\varphi$ between the two arms.
% from and phase stabilization, respectively.
}
\label{fig:setup_sagnac}
\end{figure}
Therefore, in our experiments we resorted to using an unstabilized Sagnac interferometer as shown in 
Fig.~\ref{fig:setup_sagnac}. 
% In the following, let us discuss all depicted components of the interferometer and explain their purpose. 
Although all optical components were mounted carefully to avoid birefringence as induced 
by mechanical stress, still an unwanted polarization dependent phase shift due to the total internal reflection 
at the prism remains (Goos-H\"anchen effect). This phase shift would for example rotate diagonally polarized 
light to elliptically polarized light. In order to compensate this phase shift, two YVO$_{4}$ crystals with their 
optical axis crossed were utilized (a zero-order configuration). A motorized thin piece of glass ($\approx 120\,\mu$m thick) 
was applied to set the phase difference between the two interferometer paths $a$ and $b$ to any wanted value. 
The interferometer could be balanced by a second piece of glass plate of the same thickness. Please note that the 
second glass plate was aligned such that the relative phase between the modes is compensated, i.e.,
the transformation of the input polarizing beam splitter was $|H\rangle \longrightarrow |H a\rangle$ and 
$|V\rangle \longrightarrow |V b\rangle$ for photons entering the interferometer. 
The wave plates inside the interferometer are required to analyze the 
polarization degree of freedom. In order to keep the setup as compact as possible, the quarter waveplate covers 
both spatial modes. The polarization analysis in both outputs of the interferometer enables to measure the path 
degree of freedom, i.e., allowing to distinguish between e.g. $|a+b\rangle$ and $|a-b\rangle$. The task is now to 
find angle settings for the waveplates inside and outside of the interferometer such that a tomographically complete 
set of projection measurements is obtained. Let us therefore review the (unitary) transformations that are induced 
by the respective waveplates. For the half waveplate the transformation is
\begin{equation}
 U_{\rm HWP}(\theta) =
 \begin{pmatrix}
  \cos(2 \theta) & \sin(2 \theta) \\
  \sin(2 \theta) & -\cos(2 \theta) 
  \end{pmatrix}
\end{equation}
and correspondingly for the quarter waveplate
\begin{align}
 &&\hspace{-0.75cm}U_{\rm QWP}(\theta) = \nonumber  \\
 &=&\begin{pmatrix}
  \cos(\theta)^2 - {\rm i}\sin(\theta)^2 & (1+{\rm i})\cos(\theta)\sin(\theta) \\
  (1+{\rm i})\cos(\theta)\sin(\theta) & -{\rm i}\cos(\theta)^2 + \sin(\theta)^2 
  \end{pmatrix}
\end{align}
with $|H\rangle = (1,0)^{T}$ and $|V\rangle = (0,1)^{T}$. As common to most multiqubit experiments, we choose to measure 
in the eigenbases of all combinations of local Pauli bases. In order to make the general procedure for finding 
the angle settings more illustrative, let us discuss the ${\cal M}_{31}$ basis as an example. Measuring in this basis 
means that projections onto its eigenvectors have to be performed, i.e., on 
$|H(a+b)\rangle$, $|H(a-b)\rangle$, $|V(a+b)\rangle$ and $|V(a-b)\rangle$. If one wants to project onto $|H(a+b)\rangle$ 
for example, the waveplates inside the interferometer have to transform the state just behind the polarizing input 
beam splitter such that $|Ha\rangle \longrightarrow {\rm e}^{{\rm i} \xi_a}|Ha\rangle$ and $|Hb\rangle \longrightarrow {\rm e}^{{\rm i} \xi_b}|Vb\rangle$ 
with respective phases $\xi_a$ and $\xi_b$. 
One possible choice would be
\begin{align}
 |Ha\rangle &\longrightarrow& U_{\rm QWP}\left(0\right)U_{\rm HWP}\left(0\right)|Ha\rangle = |Ha\rangle, \\
 |Hb\rangle &\longrightarrow& U_{\rm QWP}\left(0\right)U_{\rm HWP}\left(-\frac{\pi}{4}\right)|Hb\rangle = {\rm i}|Vb\rangle,
\end{align}
where the identity operation acting on the spatial mode is omitted. 
Then, the detection of a right circular polarized photon by the polarization analysis ${\rm PA}_a$ corresponds to a successful 
projection on $|H(a+b)\rangle$. On the other hand a left circular polarized photon in ${\rm PA}_a$ would correspond to 
$|H(a-b)\rangle$. Consequently, the polarization analysis ${\rm PA}_b$ in the other output of the interferometer 
allows for projection on $|V(a+b)\rangle$ and $|V(a-b)\rangle$. 
Please note that one has to trigger on coincidence counts then, i.e., one photon from each interferometer has to be detected. 
For four qubits, i.e. two interferometers, this scheme then yields $3^4=81$ different measurement settings, where in each setting $2^4=16$ projection measurements 
are performed. In our experiment we used fiber coupled single photon counting modules (SPCM from Perkin Elmer) that were 
connected to a coincidence electronic with a coincidence window of $10\,$ns. All in all $81\times16=1296$ different projectors 
were measured and a tomographically overcomplete set of data is obtained, which is processed with the method described in \cite{PhDNikolai}.
The angles for all the measurement settings can be seen in Tab.~\ref{tab:settingsME}. It has to be noted that the given angles are
not the only possible choice to obtain a tomographically (over-)complete set of projectors.
\onecolumngrid
\begin{center}
\begin{table*}[!ht]
\caption{The angles of the waveplates of the interferometer ($\text{HWP}_{\text{IF,1}}$, $\text{QWP}_{\text{IF}}$, 
$\text{HWP}_{\text{IF,2}}$) and of the two polarization analyses ($\text{HWP}_{\text{A}}$, $\text{QWP}_{\text{A}}$; 
$\text{HWP}_{\text{B}}$, $\text{QWP}_{\text{B}}$) to perform the given projections. `HA', `VA', `HB', `VB' denote the 
detectors for the transmitted (`H') and reflected light (`V') of the PBS of the polarization analysis in the output 
modes A and B, respectively. For example, an event of the detector `HA' while measuring in the basis ${\cal M}_1\otimes{\cal M}_1$  
corresponds to a successful projection onto the state $|P(a+b)\rangle$. Please note that all angles are referenced with
respect to mode $a$ which means that for calculating the transformations induced for light in mode $b$ a minus sign
has to be added.}
\vspace{0.3cm}
\begin{tabular}[b]{c |c c c| c c c c |c c c c} 
\hline 
Basis & \multicolumn{3}{|c|}{Interferometer} & \multicolumn{4}{|c|}{Polarisation analysis} & \multicolumn{4}{|c}{Projectors} \\
 \hline \hline 
 	& $\text{HWP}_{\text{IF,1}}$ 		& $\text{QWP}_{\text{IF}}$ 		& $\text{HWP}_{\text{IF,2}}$ 		& $\text{HWP}_{\text{A}}$ 		& $\text{QWP}_{\text{A}}$ 		& $\text{HWP}_{\text{B}}$ 		& $\text{QWP}_{\text{B}}$ 		& HA & VA & HB & VB \\ \hline
${\cal M}_1\otimes{\cal M}_1$ (${\cal M}_{11}$) 	& $\frac{\pi}{8}$ 	& $0$ 		& $\frac{\pi}{8}$	& $0$			& $\frac{\pi}{4}$	& $0$			& $-\frac{\pi}{4}$	& P(a+b) & P(a-b) & M(a+b) & M(a-b) \\
${\cal M}_1\otimes{\cal M}_2$ (${\cal M}_{12}$)	& $\frac{\pi}{8}$ 	& $0$ 		& $\frac{\pi}{8}$	& $-\frac{\pi}{8}$	& $0$			& $-\frac{\pi}{8}$	& $0$			& P(a+ib) & P(a-ib) & M(a+ib) & M(a-ib) \\
${\cal M}_1\otimes{\cal M}_3$ (${\cal M}_{13}$) 	& $\frac{\pi}{8}$ 	& $0$ 		& $\frac{\pi}{8}$	& $0$			& $0$			& $\frac{\pi}{4}$	& $0$			& Pa & Pb & Ma & Mb \\
${\cal M}_2\otimes{\cal M}_1$ (${\cal M}_{21}$) 	& $0$ 	& $\frac{\pi}{4}$ 		& $0$			& $0$			& $-\frac{\pi}{4}$	& $0$			& $\frac{\pi}{4}$	& R(a+b) & R(a-b) & L(a+b) & L(a-b) \\
${\cal M}_2\otimes{\cal M}_2$ (${\cal M}_{22}$) 	& $0$ 	& $\frac{\pi}{4}$ 		& $0$			& $\frac{\pi}{8}$	& $0$			& $\frac{\pi}{8}$	& $0$			& R(a+ib) & R(a-ib) & L(a+ib) & L(a-ib) \\
${\cal M}_2\otimes{\cal M}_3$ (${\cal M}_{23}$) 	& $0$ 	& $\frac{\pi}{4}$ 		& $0$			& $0$			& $0$			& $\frac{\pi}{4}$	& $0$			& Ra & Rb & La & Lb \\
${\cal M}_3\otimes{\cal M}_1$ (${\cal M}_{31}$) 	& $0$ 	& $0$ 		& $\frac{\pi}{4}$			& $0$			& $\frac{\pi}{4}$	& $0$			& $-\frac{\pi}{4}$	& H(a+b) & H(a-b) & V(a+b) & V(a-b) \\
${\cal M}_3\otimes{\cal M}_2$ (${\cal M}_{32}$) 	& $0$ 	& $0$ 		& $\frac{\pi}{4}$			& $-\frac{\pi}{8}$	& $0$			& $-\frac{\pi}{8}$	& $0$			& H(a+ib) & H(a-ib) & V(a+ib) & V(a-ib) \\
${\cal M}_3\otimes{\cal M}_3$ (${\cal M}_{33}$) 	& $0$ 	& $0$ 		& $\frac{\pi}{4}$			& $0$			& $0$			& $\frac{\pi}{4}$	& $0$			& Ha & Hb & Va & Vb 
\vspace{0.0cm}\\
\hline \hline
\end{tabular}
\label{tab:settingsME}
\vspace{1cm}
\end{table*}
\end{center}

\FloatBarrier
\twocolumngrid
\section*{SM\,4: Results}
\begin{figure}[!ht]
\includegraphics[width=0.80\textwidth]{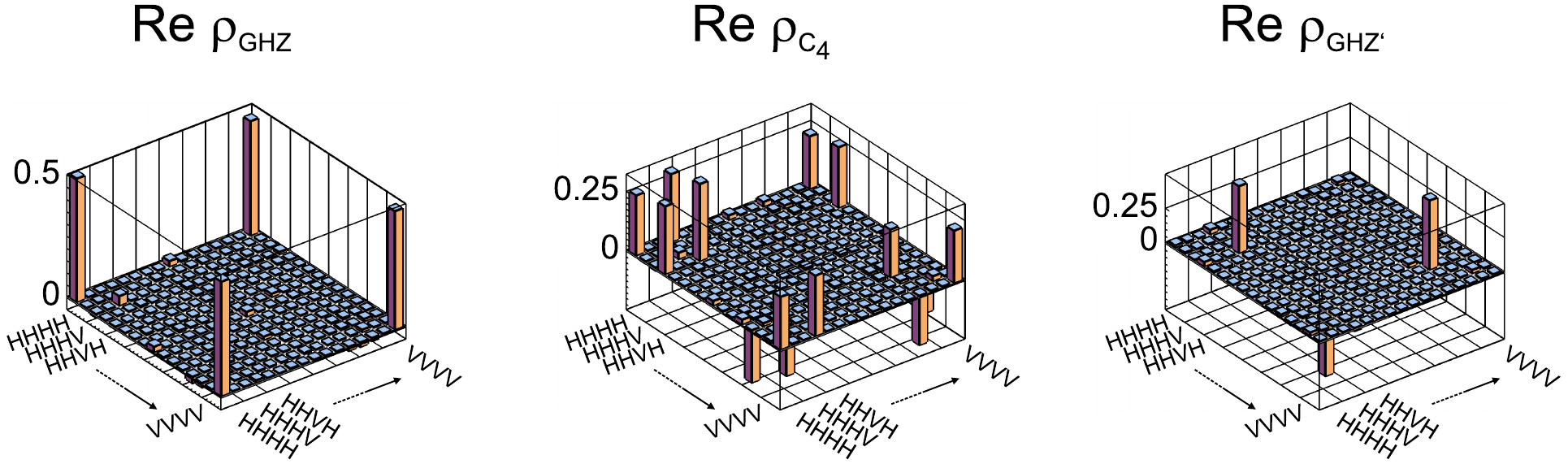}
\caption{Real parts of the experimental density matrices of (a) the GHZ state ($\theta=0$, $\phi=\pi$), (b) the cluster state ($\theta=\pi/8$, $\phi=\pi$), 
and (c) the GHZ$^\prime$ state ($\theta=\pi/4$, $\phi=\pi$).}
\label{fig:density_all}
\end{figure}

We prepared and characterized $13$ states belonging to the states given by $|\Psi\left(\theta,\phi\right)\rangle$ in the main text, including the GHZ and Cluster state. 
For all states, we carried out full quantum state tomography. 
Fig.~\ref{fig:density_all} shows the experimental density matrices of the GHZ state, the cluster state, and for another GHZ-type state $|{\rm GHZ}^\prime \rangle = \left(|0011\rangle-|1100\rangle\right)/\sqrt{2}$. 
From the density matrices, the fidelity with the theoretically expected  states could be inferred, see Tab.~\ref{tab:statesResults}. 
The fidelities of the prepared states compared with the respective target state were above $95.8\%$ in all cases. 
The correlation values used for the witnesses are listed in Tab.~\ref{tab:expCorr} for the GHZ and Cluster state.

\onecolumngrid
\begin{center}
\begin{table}[bthp]
\caption{Experimentally determined correlation values for the GHZ and Cluster state, which are used to calculate the values of the two measurement witnesses. For the GHZ state, the measurement settings  ${\cal M}_{1221}$ and ${\cal M}_{3333}$ were used and for the Cluster state ${\cal M}_{1133}$ and ${\cal M}_{3311}$.
}
\vspace{0.2cm}
\begin{tabular}[b]{|c |r |c| c|r |} 
\hline \hline 
\multicolumn{2}{|c|}{GHZ} & & \multicolumn{2}{|c|}{Cluster} \\
\hline
$T_{3333}$ & $0.982 \pm 0.003$ &&  $T_{3300}$ & $0.987 \pm 0.002$ \\
$T_{3300}$ & $0.993 \pm 0.002$ &&  $T_{3011}$ & $0.986 \pm 0.003$ \\ 
$T_{0033}$ & $0.988 \pm 0.002$ &&  $T_{0311}$ & $0.974 \pm 0.003$ \\ 
$T_{3003}$ & $0.963 \pm 0.004$ &&  $T_{1130}$ & $-0.945 \pm 0.006$ \\ 
$T_{0330}$ & $0.969 \pm 0.004$ &&  $T_{1103}$ & $-0.934 \pm 0.006$ \\
$T_{3030}$ & $0.972 \pm 0.004$ &&  $T_{0033}$ & $0.989 \pm 0.002$ \\
$T_{0303}$ & $0.960 \pm 0.005$ && & \\
$T_{1221}$ & $-0.925 \pm 0.006$ && & \\
\hline \hline
\end{tabular}
\label{tab:expCorr}
\end{table}
\end{center}
\twocolumngrid

For all prepared states, at least one of the two combined witnesses ${\cal W}^{\mathrm{GHZ}}$ and ${\cal W}^{\mathrm{C}_4}$, which both could be determined from two measurement settings only, lies above the respective threshold ($7/11$ for the GHZ criterion or $2/3$ for the cluster criterion). 
Therefore, genuine four-partite entanglement could be proven for all considered states. 
\onecolumngrid
\begin{center}
\begin{table*}[!ht]
\caption{Characterization of $13$ states given by $|\Psi\left(\theta,\phi\right)\rangle$ in the main text with $\phi=\pi$.
The fidelities with the respective target states were determined from the experimental density matrices as obtained via quantum state tomography.
The values for the entanglement criteria ${\cal W}^{\mathrm{GHZ}}$ and ${\cal W}^{{\cal C}_4}$ as presented in the main text, however, were inferred from two measurement settings only. 
For all prepared states, genuine four-partite entanglement can be proved by at least one of the two criteria. Successful entanglement detection of the respective criterion 
is indicated by bold letters.}
\vspace{0.3cm}
\begin{tabular}[b]{c c c c c c c c c c} 
\hline \hline 
Name 	& $\theta$ 		&~~& Fidelity ${\cal F}$ &~~~& ${\cal W}^{\mathrm{GHZ}} \leq_{\rm BISEP} \frac{7}{11}$&~~~& ${\cal W}^{{\cal C}_4} \leq_{\rm BISEP} \frac{2}{3}$ 
\vspace{0.05cm}\\ \hline 
GHZ 	& $0$ 			& & $0.958\pm0.004$ 	& & $\boldsymbol{0.916\pm0.005}$ & & $0.333\pm0.002$  \\
	& $\frac{\pi}{48}$ 	& & $0.959\pm0.004$ 	& & $\boldsymbol{0.894\pm0.006}$ & & $0.387\pm0.005$  \\
	& $\frac{\pi}{24}$	& & $0.958\pm0.004$ 	& & $\boldsymbol{0.828\pm0.007}$ & & $0.509\pm0.007$  \\
	& $\frac{\pi}{16}$ 	& & $0.965\pm0.003$ 	& & $\boldsymbol{0.740\pm0.007}$ & & $\boldsymbol{0.685\pm0.007}$  \\
	& $\frac{\pi}{12}$ 	& & $0.963\pm0.003$ 	& & $\boldsymbol{0.644\pm0.006}$ & & $\boldsymbol{0.835\pm0.006}$  \\
	& $\frac{5\pi}{48}$ 	& & $0.963\pm0.003$ 	& & $0.603\pm0.004$ & & $\boldsymbol{0.918\pm0.005}$  \\
Cluster & $\frac{\pi}{8}$ 	& & $0.962\pm0.003$ 	& & $0.590\pm0.004$ & & $\boldsymbol{0.940\pm0.004}$  \\
	& $\frac{7\pi}{48}$ 	& & $0.959\pm0.004$ 	& & $0.608\pm0.006$ & & $\boldsymbol{0.886\pm0.007}$  \\
	& $\frac{\pi}{6}$ 	& & $0.959\pm0.004$ 	& & $\boldsymbol{0.679\pm0.008}$ & & $\boldsymbol{0.761\pm0.008}$  \\
	& $\frac{3\pi}{16}$ 	& & $0.958\pm0.003$ 	& & $\boldsymbol{0.746\pm0.007}$ & & $0.640\pm0.008$  \\
	& $\frac{5\pi}{24}$ 	& & $0.960\pm0.004$ 	& & $\boldsymbol{0.820\pm0.007}$ & & $0.497\pm0.007$  \\
	& $\frac{11\pi}{48}$ 	& & $0.963\pm0.004$ 	& & $\boldsymbol{0.890\pm0.006}$ & & $0.375\pm0.004$  \\
GHZ$^\prime$ & $\frac{\pi}{4}$ 	& & $0.967\pm0.004$ 	& & $\boldsymbol{0.927\pm0.005}$ & & $0.330\pm0.002$  \\
\hline \hline
\end{tabular}
\label{tab:statesResults}
\vspace{1cm}
\end{table*}
\end{center}

\FloatBarrier

\end{document}